\documentclass[prl,onecolumn,superscriptaddress,preprintnumbers,amsmath,amssymb]{revtex4}
\usepackage{graphicx}
\usepackage{amsmath}
\usepackage{latexsym}
\usepackage{dcolumn}
\usepackage{bm}% bold math
\usepackage{float}
\usepackage{graphicx,here}

\emergencystretch=\maxdimen
\begin{document}

\title{Dirac fermions in borophene}

\author{Baojie Feng}
\affiliation{Institute for Solid State Physics, The University of Tokyo, Kashiwa, Chiba 277-8581, Japan}
\author{Osamu Sugino}
\affiliation{Institute for Solid State Physics, The University of Tokyo, Kashiwa, Chiba 277-8581, Japan}
\author{Ro-Ya Liu}
\affiliation{Institute for Solid State Physics, The University of Tokyo, Kashiwa, Chiba 277-8581, Japan}
\author{Jin Zhang}
\affiliation{Institute of Physics, Chinese Academy of Sciences, Beijing 100190, China}
\author{Ryu Yukawa}
\affiliation{Institute of Materials Structure Science, High Energy Accelerator Research Organization (KEK), Tsukuba, Ibaraki 305-0801, Japan}
\author{Mitsuaki Kawamura}
\affiliation{Institute for Solid State Physics, The University of Tokyo, Kashiwa, Chiba 277-8581, Japan}
\author{Takushi Iimori}
\affiliation{Institute for Solid State Physics, The University of Tokyo, Kashiwa, Chiba 277-8581, Japan}
\author{Howon Kim}
\affiliation{Institute for Solid State Physics, The University of Tokyo, Kashiwa, Chiba 277-8581, Japan}
\author{Yukio Hasegawa}
\affiliation{Institute for Solid State Physics, The University of Tokyo, Kashiwa, Chiba 277-8581, Japan}
\author{Hui Li}
\affiliation{Institute of Physics, Chinese Academy of Sciences, Beijing 100190, China}
\author{Lan Chen}
\affiliation{Institute of Physics, Chinese Academy of Sciences, Beijing 100190, China}
\author{Kehui Wu}
\affiliation{Institute of Physics, Chinese Academy of Sciences, Beijing 100190, China}
\affiliation{Collaborative Innovation Center of Quantum Matter, Beijing 100871, China}
\author{Hiroshi Kumigashira}
\affiliation{Institute of Materials Structure Science, High Energy Accelerator Research Organization (KEK), Tsukuba, Ibaraki 305-0801, Japan}
\author{Fumio Komori}
\affiliation{Institute for Solid State Physics, The University of Tokyo, Kashiwa, Chiba 277-8581, Japan}
\author{Tai-Chang Chiang}
\affiliation{Department of Physics, University of Illinois, Urbana, IL 61801, USA}
\affiliation{Institute for Solid State Physics, The University of Tokyo, Kashiwa, Chiba 277-8581, Japan}
\author{Sheng Meng}
\affiliation{Institute of Physics, Chinese Academy of Sciences, Beijing 100190, China}
\affiliation{Collaborative Innovation Center of Quantum Matter, Beijing 100871, China}
\author{Iwao Matsuda}
\thanks{imatsuda@issp.u-tokyo.ac.jp}
\affiliation{Institute for Solid State Physics, The University of Tokyo, Kashiwa, Chiba 277-8581, Japan}

\date{\today}

%% maintext

\begin{abstract}
Honeycomb structures of group IV elements can host massless Dirac fermions with non-trivial Berry phases. Their potential for electronic applications has attracted great interest and spurred a broad search for new Dirac materials especially in monolayer structures. We present a detailed investigation of the $\beta_{12}$ boron sheet, which is a borophene structure that can form spontaneously on a Ag(111) surface. Our tight-binding analysis revealed that the lattice of the $\beta_{12}$-sheet could be decomposed into two triangular sublattices in a way similar to that for a honeycomb lattice, thereby hosting Dirac cones. Furthermore, each Dirac cone could be split by introducing periodic perturbations representing overlayer-substrate interactions. These unusual electronic structures were confirmed by angle-resolved photoemission spectroscopy and validated by first-principles calculations. Our results suggest monolayer boron as a new platform for realizing novel high-speed low-dissipation devices.
\end{abstract}

\maketitle

Nontrivial lattice structures of solids involving more than one atom per lattice site can host novel properties and behaviors; hence, the discovery and design of new structure-property combinations are at the forefront of materials science. A celebrated, but particularly simple, example is the two-dimensional honeycomb lattice with just two atoms per unit cell (Fig. 1a), such as graphene\cite{Geim2007,Neto2009}, silicene\cite{Houssa2015,Zhuang2015,Zhao2016}, germanene\cite{LiLF2014,Derivaz2015} and stanene\cite{Zhu2015}. These materials can host Dirac cones that give rise to rich physical properties\cite{Neto2009,Feng2013,Ezawa2012,XuY2013}. Recent theoretical investigations for new Dirac materials in simple two-dimensional structures have attracted great attention\cite{Malko2012,ZhangLZ2015,WangJ2015,Miert2016,ZhouXF2014}, whereas experimental observations of Dirac cones beyond the honeycomb structure are still rare.

A promising route to realize novel two-dimensional materials is by tailoring or modifying the honeycomb lattice. An example is a monolayer boron sheet (i.e., borophene), which is realized by introducing periodic boron atoms in a honeycomb-like lattice. As boron has one less electron than carbon, the honeycomb structure is unstable but the introduction of additional boron atoms in the honeycomb lattice can stabilize the structures by balancing out the two- and multi-center bonds\cite{Evans2005,Tang2010,Tang2007}. Depending on the arrangements of the extra boron atoms, various monolayer-boron structures have been proposed, such as the $\alpha$-sheet, $\beta$-sheet, etc\cite{Tang2010,Tang2007,WuX2012,Penev2012,LiuH2013}. Recently, several monolayer boron phases have been experimentally realized on Ag(111)\cite{Mannix2015,Feng2016,Zhang2016,Feng2016'}. For example, Mannix {\it et al.} reported a stable striped phase and a metastable homogeneous phase\cite{Mannix2015}. The striped phase was proposed to be a complete triangular lattice with anisotropic, out-of-plane buckling. In another study, a similar striped phase with a different rotation angle has been observed\cite{Feng2016}. This phase, named a $\beta_{12}$-sheet (Fig. 1b), has an essentially flat structure and interacts weakly with the Ag(111) substrate\cite{Feng2016,Feng2016',Zhang2015,LiuY2013}. However, the experimental investigations on the electronic properties of monolayer boron are still rare.

In this letter, we present a combined theoretical and experimental investigations on the $\beta_{12}$ boron sheet. Our tight-binding analysis reveals that the lattice $\beta_{12}$-sheet can be decomposed into two triangular sublattices, analogous to the honeycomb lattice, and thus hosts Dirac cones. Moreover, each Dirac cone can be split by introducing periodic perturbations representing the moir\'{e} pattern observed by a scanning tunneling microscope. These intriguing electronic structures have been confirmed by angle-resolved photoemission spectroscopy (ARPES) measurements and first-principles calculations. Our results have experimentally confirmed the first monolayer Dirac materials beyond the honeycomb structure and have validated a novel approach to split the Dirac cones by periodic perturbations. Moreover, these results suggest monolayer boron as a promising material to realize high-speed, low-dissipation nanodevices.

In graphene, the $\pi$ bands near the Fermi level (E$\rm_F$) derived from the p$_z$ orbital form the Dirac cones at the K points (Fig. 1a)\cite{Neto2009}. The s, p$_x$ and p$_y$ orbitals are sp$^2$ hybridized and contribute to the $\sigma$ bands which are far from E$\rm_F$. The $\beta_{12}$-sheet is also atomically flat, as is graphene, and, as confirmed later by our experiments and first-principles calculations, the bands near E$\rm_F$ are also derived from the p$_z$ orbital. Interestingly, a simple tight-binding (TB) model, considering only the p$_z$ orbital for a freestanding $\beta_{12}$-sheet, shows the existence of Dirac cones centered at ($\pm2\pi/3a$, $0$) in the first Brillouin zone (BZ), as illustrated in Fig. 1b.

\begin{figure}[htbp]
\includegraphics[width=8.5 cm]{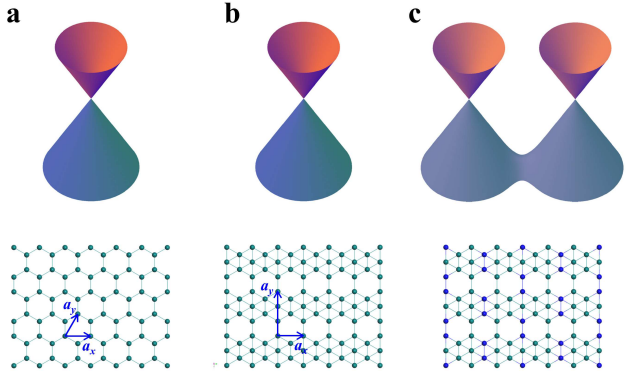}
\caption {{\bf Schematic drawing of the Dirac cones and lattices.} (a) Honeycomb structure. (b) $\beta_{12}$-sheet. (c) The $\beta_{12}$-sheet with a 3$\times$1 perturbation. The blue and green balls indicate the boron atoms with different on-site energies in our TB analysis. The top and bottom panels are the band structures and atomic structures, respectively. The basic vectors of the primitive unit cell are indicated by the blue arrows.}
\end{figure}

For a detailed understanding of the electronic structure of the system, we present the wave function for each boron atom in Fig. 2a. Our TB analysis showed that the wave function at E$\rm_F$ has a vanishing amplitude at site {\it c}, owing to phase cancellation at the six-fold coordinated boron atoms. The wave function originates instead from the atoms at sites {\it a}, {\it b}, {\it d} and {\it e}, which can be decomposed into two sublattices (Figs. 2a and 2b). As a result, the equivalent structure of the $\beta_{12}$-sheet is a honeycomb lattice, as shown in Fig. 2c. As with graphene, this honeycomb lattice gives rise to a Dirac cone at each $\rm\overline{K}$ point of the BZ. These Dirac cones are folded to ($\pm2\pi/3a$, $0$), as illustrated in Fig. 2d, because the B atoms at site {\it c} alter the shape and size of the BZ. The band structure from our TB calculations is shown in Fig. 2e, where the two Dirac cones in the $\Gamma$-X axis are indicated by black arrows. For further confirmation, we also performed first-principles calculations for the freestanding $\beta_{12}$-sheet, and the Dirac cones at ($\pm2\pi/3a$, $0$) were reproduced (Fig. 2f). The Dirac points were located at approximately 2 eV above the Fermi level, in agreement with previous reports\cite{Penev2016}. The upward shift of the Dirac cones might originate from the electron deficiency of boron. It should be noted that the energy position of the Dirac points can be varied after being placed on a metal substrate to compensate for the electron deficiency (see Supplementary Materials for details). Similar Dirac cone states in a rectangular lattice have also recently been proposed in a graphene superlattice\cite{Jia2016}.

When the $\beta_{12}$-sheet is placed on a Ag(111) substrate, a long-range modulation arising from the lattice mismatch gives rise to a moir\'{e} pattern, as shown in Fig. S4. As the interaction of the boron layer and Ag(111) substrate is weak, the $\beta_{12}$-sheet remains largely intact and the moir\'{e} pattern can be explained by a modulated charge distribution on the surface\cite{Feng2016}. The long-range modulation yields an electronic perturbation; in our TB model, we simulate this effect by varying the on-site energy over a superlattice period of {\it na$_x$}$\times${\it ma$_y$}, where {\it a$_x$}$\times${\it a$_y$} is the original unit cell (Fig. 1b). The Dirac cones of the superlattice are folded onto the $\Gamma$ point when {\it n} is a multiple of three, and are further split into pairs in the $\Gamma$-Y direction when the sublattice symmetry is broken while retaining the inversion symmetry (Figs. 1c and 2g). The Dirac cones will split in the $\Gamma$-X direction when the inversion symmetry is also broken (see the Supplementary Materials for details). The splitting of the Dirac cones has also been confirmed by our first-principles calculations considering the periodic perturbation (Fig. S2b). From Fig. 2g, the split Dirac cones are non-concentric which is different from the Rashba-type splitting of the Dirac cones in graphene\cite{Varykhalov2008,Marchenko2012}.

\begin{figure*}[htbp]
\includegraphics[width=16 cm]{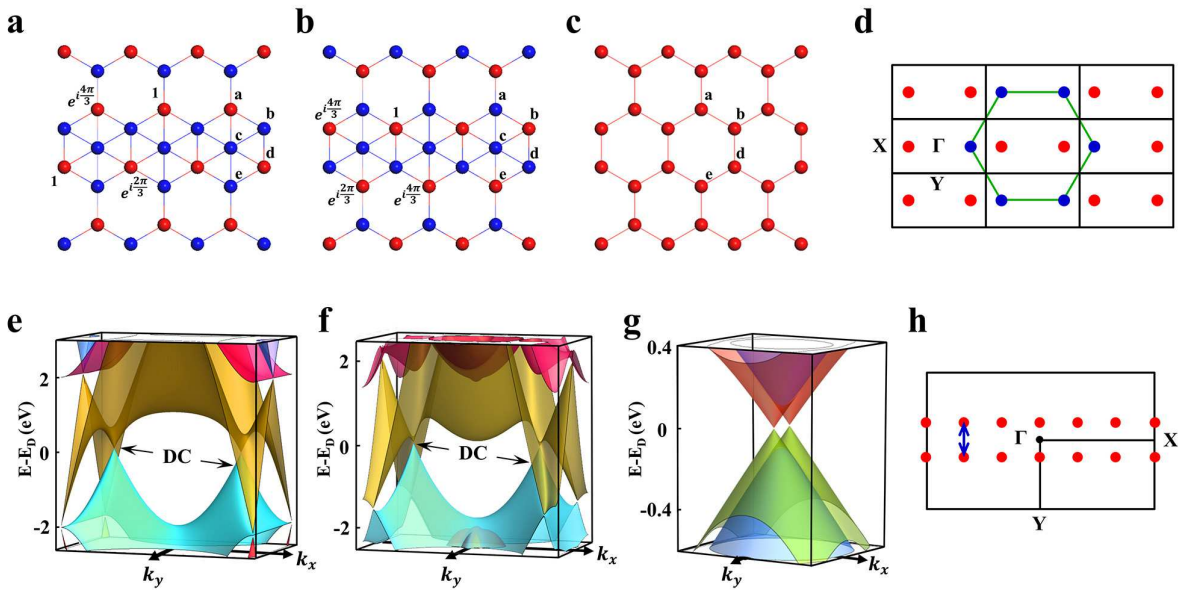}
\caption{{\bf TB model and first-principles calculations of the $\beta_{12}$-sheet.} (a) and (b) The wave function of each boron atom, as indicated by the amplitude near the red balls; the blue balls represent boron atoms with a vanishing amplitude. The boron atoms at site {\it c} always have a vanishing amplitude. The red balls are equivalent to the sublattices of the honeycomb lattice in (c). (d) Schematic drawing of the band folding process. The green hexagon and black rectangles indicate the BZ of the equivalent honeycomb lattice and the $\beta_{12}$-sheet, respectively. The blue dots indicate the original Dirac cones from the honeycomb lattice; the red dots indicate the folded Dirac cones. (e) and (f) Band structures of free-standing $\beta_{12}$-sheet from the TB model and first-principles calculations, respectively. E$\rm_D$ in the figure corresponds to the Dirac point, which is approximately 2 eV above the Fermi level from our first-principles calculations. The black arrows indicate the Dirac cones. (g) TB band structures of the $\beta_{12}$-sheet under a modulated potential. The Dirac cone is split in the $\Gamma$-Y direction. (h) Schematic drawing of the folding and splitting (blue arrow) of the Dirac cones.}
\end{figure*}

\begin{figure*}[htbp]
\includegraphics[width=16 cm]{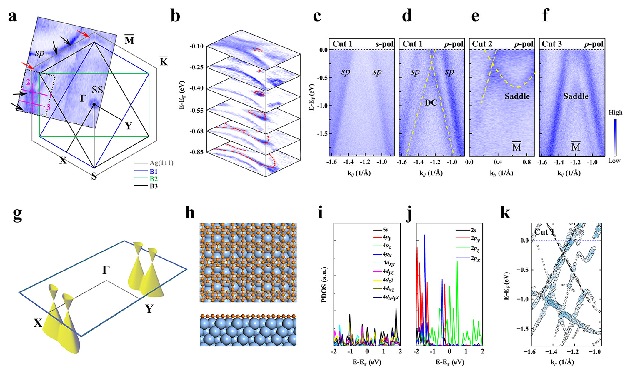}
\caption{{\bf Band structures of the $\beta_{12}$-sheet on Ag(111).} (a) The Fermi surface of the $\beta_{12}$-sheet on Ag(111). The black, green, and blue rectangles indicate the BZ of three equivalent domains; the grey hexagon indicates the BZ of Ag(111). The black and red arrows indicate the bands of the boron layer. The surface state (SS) and bulk {\it sp} band of Ag(111) are also observed because the coverage of boron is less than 1 ML. The pink lines indicate cuts 1-3 where the ARPES intensity plots in (c)-(f) were measured. (b) CECs derived from the second-derivative EDCs measured in the black dotted rectangle in (a). E$\rm_F$ in the figure corresponds to the Fermi level. All the data in (a) and (b) were measured with {\it p} polarized light. (c) ARPES intensity plot measured along cut 1 with {\it s} polarized light. (d)-(f) ARPES intensity plots measured with {\it p} polarized light along cut 1 to cut 3, respectively. The yellow dashed lines indicate the Dirac cones (DC). All the ARPES data in (a)-(f) were measured with a photon energy of 80 eV. (g) Schematic drawing of the Dirac cones according to our experimental results. (h) Relaxed structure model of the $\beta_{12}$-sheet on Ag(111) from our first-principles calculations. The orange and blue balls indicate the B and Ag atoms, respectively. (i) and (j) Calculated PDOS of B atoms and Ag atoms, respectively. (k) Calculated band structure along cut 1.}
\end{figure*}

To confirm these intriguing properties of the $\beta_{12}$-sheet, we have performed high-resolution angle-resolved photoemission spectroscopy (ARPES) to directly measure its band structure. The sample was prepared by evaporating pure boron onto a Ag(111) substrate (see Supplementary Materials for details). From LEED measurements (Fig. S4a), we found that there is only one phase, the $\beta_{12}$-sheet. As the $\beta_{12}$-sheet has a rectangular structure, different from the hexagonal structure of Ag(111), there exist domains with three equivalent orientations related by 120$^{\circ}$ rotations. A schematic drawing of the BZ of Ag(111) with the three domain orientations is shown in Fig. 3a, together with the measured Fermi surface. Because the coverage of boron was less than one monolayer in the experiments, there were some areas of bare Ag(111) surface. As a result, the Shockley surface state and bulk $\textit{sp}$ band of Ag(111) were clearly observed, as indicated by ``{\it SS}" and ``{\it sp}" in Fig. 3a. The band structure from the boron layer shows one Fermi pocket centered at the $\rm\overline{S}$ point of the $\beta_{12}$-sheet and a pair of Fermi pockets centered at the $\rm\overline{M}$ point of Ag(111), as indicated by the red and black arrows, respectively. The bands derived from the boron layer do not disperse with an increasing photon energy (Fig. S5), which is in agreement with its two-dimensional characteristic.

The pair of Fermi pockets centered at the $\rm\overline{M}$ point of Ag(111) is associated with Dirac cones, in agreement with the general picture based on our calculations. In Fig. 3b, we show constant energy contours (CECs) at different binding energies (E$_B$). With increasing binding energies, the Fermi pockets first shrink in size and then become points at E$_B$ = -0.25 eV. Further increase of the binding energy leads to a pair of closed contours which touch each other at E$_B$ = -0.68 eV. The pair of closed contours merge into one contour at higher binding energies.

The band structure measured along typical cuts in the momentum space (the purple lines in Fig. 3a) is shown in Figs. 3c-3f. The measurements of cut 1 using {\it p} polarized light (Fig. 3d) reveal a Dirac cone as well as the bulk {\it sp} band of Ag(111). The Dirac point is located at approximately 0.25 eV below the Fermi level, in agreement with the evolution of the CECs in Fig. 3b. The linear dispersing bands extend to as deep as 2 eV. Within our experimental resolution, there is no obvious energy gap at the Dirac point; thus the quasiparticles are massless Dirac fermions. The Fermi velocities determined from Fig. 3d are approximately 6.1 eV$\cdot${\AA} and 7.0 eV$\cdot${\AA} for the left and right branches of the Dirac cone, respectively, which are close to the Fermi velocity of graphene ($\sim$6.6 eV$\cdot${\AA}). The slight difference of the Fermi velocity between the two branches originates from the anisotropy of the Dirac cones, in agreement with the CECs in Figs. 3(b) and S5. The neighboring bulk {\it sp} band of Ag(111) is clearly separated from the Dirac cone with no signs of hybridization, which indicates a weak interaction between the $\beta_{12}$-sheet and the Ag(111) substrate, in agreement with previous work\cite{Zhang2015}. In Fig. 3e, we show the band structure along the $\rm\overline{K}$-$\rm\overline{M}$-$\rm\overline{K}$ direction; a pair of Dirac cones can be identified (indicated by the yellow dashed lines) although the one on the right side is only half-visible because of the limitation of our experimental configuration. The two cones touch each other at the $\rm\overline{M}$ point of Ag(111) at a binding energy of approximately 0.68 eV, which agrees with the evolution of the CECs discussed above. The band structure at the $\rm\overline{M}$ point of Ag(111) shows a ``V" shape along the $\rm\overline{K}$-$\rm\overline{M}$-$\rm\overline{K}$ direction (Fig. 3e) and a ``$\Lambda$" shape along the $\overline{\Gamma}$-$\overline{\rm M}$ direction (Fig. 3f); the bottom of the ``V" and the top of the ``$\Lambda$" are located at the same binding energy ($\sim$ 0.68 eV), which suggests a ``saddle" point at the $\overline{\rm M}$ point of Ag(111). Within the first BZ of the $\beta_{12}$-sheet, we observed two pairs of Dirac cones in total, as schematically illustrated in Fig. 3g.

The orbital contribution of the boron bands can be probed by switching the linear polarization of the incident light. The {\it s} polarized light primarily probes the in-plane {\it p$_x$} and {\it p$_y$} orbitals, while the {\it p} polarized light probes both the in-plane ({\it p$_x$} and {\it p$_y$}) and out-of-plane ({\it p$_z$}) orbitals. The band structures along cut 1 measured with {\it s} and {\it p} polarized light are shown in Figs. 3c and 3d, respectively. The Dirac cone was not observed with the {\it s} polarized light, leaving only the bulk {\it sp} bands of Ag(111). This means that the Dirac cones originate from the {\it p$_z$} orbital of boron. For further confirmation, we performed first-principles calculations for the B/Ag(111) system. The relaxed atomic structure shown in Fig. 3h agrees with previous work\cite{Feng2016}. The partial density of states (PDOS) of the boron and silver atoms is shown in Figs. 3i and 3j. Near E$\rm_F$, the density of states (DOS) is mainly derived from the {\it p$_z$} orbital of boron; the contributions from the {\it p$_x$} and {\it p$_y$} orbitals of boron are essentially negligible (Fig. 3i). Likewise, the contributions from Ag orbitals are much smaller compared with those from the {\it p$_z$} orbital of boron (Fig. 3j). We conclude that the Dirac cones are predominantly derived from the {\it p$_z$} orbital of boron, with little hybridization with the Ag substrate states. This observation validates our TB analysis in terms of the boron {\it p$_z$} orbital only.

Although the contributions from the Ag atoms is much smaller than the {\it p$_z$} orbital of boron, there are still considerable contributions from the 5{\it s}, 4{\it d$_{z^2}$}, 4{\it d$_{xz}$} and 4{\it d$_{yz}$} orbitals of Ag atoms over much of the valence band range. These orbitals have large out-of-plane components and could potentially hybridize with the {\it p$_z$} orbital of boron, which can explain the origin of the weak interaction between the $\beta_{12}$-sheet and the Ag(111) substrate. This interaction might energetically shift the bands of the free-standing $\beta_{12}$-sheet, moving the Dirac points below the Fermi level. Another important consequence of this interaction is the appearance of the moir\'{e} pattern. From Fig. S4b, the period of the moir\'{e} pattern is approximately 5.5{\it a$_{x}$}, which is approximately two times the period of the perturbation in our TB model (Fig. 1c). This observation validates our qualitative explanation for the splitting of the Dirac cones. Alternatively, the splitting of the Dirac cones could be interpreted in terms of a uniaxial strain in the $\beta_{12}$-sheet associated with the moir\'{e} pattern. The strain in the lattice could break the equivalence of bonds, inducing a splitting of the $\pi$ bands\cite{Jia2016}. The net results would be similar to those caused by a modulation of the on-site energy. On the other hand, owing to the existence of the moir\'{e} pattern, the pair of Dirac cones centered at ($\pm2\pi/3a$, $0$) of the $\beta_{12}$-sheet are folded to the $\rm\overline{M}$ point of Ag(111), in agreement with our experiments. As a further test of our explanation, first-principles calculations of B/Ag(111) also reveal the same pair of Dirac cones, as shown in Fig. 3k. The calculated Fermi velocity is approximately 3.5 eV$\cdot${\AA}, which is in the same order of magnitude as the experimental value. The difference between the theoretical and experimental results might originate from the many-body interactions, which have already been extensively studied in graphene\cite{Elias2011,Hwang2012}.

All of our results support or confirm the existence of gapless Dirac cones in the $\beta_{12}$ boron sheet grown on Ag(111). These Dirac cones are split into pairs owing to the interaction of the boron layer with the substrate. An important implication of our analysis and discussion of the underlying physics is that Dirac cone features can arise in lattices with large unit cells; such systems tend to exhibit multiple motifs and are conducive to atomic scale engineering of the structure. Our work suggests opportunities and strategies in connection with the realization of Dirac and possibly other exotic phases; it might also stimulate further investigation of the novel properties of monolayer boron, such as superconductivity\cite{Penev2016}, topological order, and high-speed electronic transport and switching.

\begin{acknowledgements}
We thank Professor X. J. Zhou for providing the Igor macro to analyze the ARPES data. This work was supported by the Synchrotron Radiation Research Organization at the University of Tokyo, the Ministry of Education, Culture, Sports, Science and Technology of Japan (Photon and Quantum Basic Research Coordinated Development Program), the JSPS grant-in-aid for specially promoting research (Grant No. 23000008), the JSPS grant-in-aid for Scientific Research (B) (Grant No. 26287061), Japan Science and Technology Agency (JST) ACT-C, the US National Science Foundation (Grant No. DMR-1305583), the MOST of China (Grants Nos. 2013CB921702, 2013CBA01601, 2016YFA0202301), the NSF of China (Grant Nos. 11322431, 1674366), and the Strategic Priority Research Program of the Chinese Academy of Sciences (Grant No. XDB07020100).
\end{acknowledgements}

\clearpage

\begin{center}

{\Large\bf Supplementary Materials for}

{\Large\bf Dirac fermions in borophene}

\end{center}

\leftline{\bf 1. First-principles and TB analysis of the split Dirac cones}

\bigskip We have investigated the band structure of the free-standing $\beta_{12}$-sheet using a mixed first-principles and tight-binding (TB) model approach. The first-principles calculation shown in Fig. 2f is based on the density functional theory (DFT) within the generalized gradient approximation (GGA). The calculation was performed without geometry optimization ($a=2.9236$ \AA \ and $b/a=\sqrt{3}$) using the program package \textsc{Quantum ESPRESSO}\cite{QE}, Elk\cite{Elk} and OpenMX\cite{OpemXM}. In the \textsc{Quantum ESPRESSO} calculation, we used the plane wave basis set with a cut-off of $50$ Ry for the valence electrons and the core electrons were treated using the ultrasoft pseudopotential \cite{USPP, pseudo}. The $k$-point mesh was set to $36\times24\times1$ in the optimization of the charge density. The resulting band structure was checked using the Elk within the Full Potential Linearized Augmented Plane Wave (FP-LAPW) scheme and the OpenMX using the double zeta basis set and the pseudopotential as implemented in the package\cite{vpsandpao}.

The first-principles calculation shows that the bands in the energy range $E_{F}$ to $E_{F}+4$ eV are almost completely composed of the 2p$_{z}$ orbital of boron. To analyze those bands, we constructed the maximally localized Wannier function (MLWF) \cite{wan90} of 2p$_{z}$ and obtained the effective Hamiltonian:%
\begin{equation}
\mathcal{H}=\sum_{i}\varepsilon_{i}c_{i}^{\dagger}c_{i}+\sum_{ij}%
^{\text{n.n.}}t_{ij}c_{i}^{\dagger}c_{j}+\sum_{ij}^{\neq\text{n.n.}}\tilde
{t}_{ij}c_{i}^{\dagger}c_{j}, \label{eqs1}%
\end{equation}
where $c_{i}^{\dagger}$ and $c_{i}$ are, respectively, the creation and the annihilation operator of 2p$_{z}$. The on-site energies $\varepsilon_{i}$\ of the constituting five sites $\vec{R}_{a}=\left(  0,3a_{y}\right)  $, $\vec{R}_{b}=\left(  \frac{\sqrt{3}}{2}a_{x},\frac{5}{2}a_{y}\right)  $, $\vec{R}_{c}=\left(  0,2a_{y}\right)  $, $\vec{R}_{d}=\left(  \frac{\sqrt{3}}{2}a_{x},\frac{3}{2}a_{y}\right)  $, $\vec{R}_{e}=\left(  0,a_{y}\right)  $ are%

\begin{equation}
\varepsilon_{a}=\varepsilon_{d}=0.196\text{ eV, }\varepsilon_{b}%
=\varepsilon_{e}=-0.058\text{ eV, }\varepsilon_{c}=-0.845\text{ eV.}
\label{eqs2}%
\end{equation}
The transfer $t_{ij}$ of the nearest neighbor (n.n.) pairs and that of the other pairs $\tilde{t}_{ij}$ are given in Table \ref{table1}. To determine the property of the MLWF effective Hamiltonian, we analyzed a simplified TB Hamiltonian consisting of the nearest neighbor transfer only:%
\begin{equation}
\mathcal{H}=-t_{0}\sum_{ij}^{\text{n.n.}}c_{i}^{\dagger}c_{j}, \label{eqs3}%
\end{equation}
where the transfer is taken to be common ($t_{0}=2$ eV).

For the primitive $\beta_{12}$-sheet, there is an eigensolution of Eq.\ (\ref{eqs3})\ that crosses the zero energy at $T_{1}=\left(  \frac{2}{3}\frac{\pi}{a_{x}},0\right)  $ in the BZ and another solution that crosses the
zero energy at $\bar{T}_{1}=\left(  -\frac{2}{3}\frac{\pi}{a_{x}},0\right)  $. The corresponding Bloch wave functions, $\varphi_{T_{1}}$ and $\varphi_{\bar{T}_{1}\text{,}}$ are
\begin{equation}
\varphi_{T_{1}}^{\alpha}\left(  x+ma_{x},y+na_{y}\right)  =\left\{
\begin{array}
[c]{c}%
\omega^{m}\left(  \phi_{a}\left(  x,y\right)  +\omega^{2}\phi_{d}\left(
x,y\right)  \right)  \text{ when }\alpha=1\\
\omega^{m}\left(  \phi_{e}\left(  x,y\right)  +\omega^{2}\phi_{b}\left(
x,y\right)  \right)  \text{ when }\alpha=2
\end{array}
\right.  \label{eqs5a}%
\end{equation}
and
\begin{equation}
\varphi_{\bar{T}_{1}}^{\alpha}\left(  x+ma_{x},y+na_{y}\right)  =\left\{
\begin{array}
[c]{c}%
\omega^{-m}\left(  \phi_{a}\left(  x,y\right)  +\omega^{-2}\phi_{d}\left(
x,y\right)  \right)  \text{ when }\alpha=1\\
\omega^{-m}\left(  \phi_{e}\left(  x,y\right)  +\omega^{-2}\phi_{b}\left(
x,y\right)  \right)  \text{ when }\alpha=2
\end{array}
\right.  , \label{eqs5b}%
\end{equation}
where $\phi_{p}$ is the atomic orbital, $\omega$ is a complex number satisfying $\omega^{3}=-1$, and $m,n$ are integers. Importantly, $\varphi_{T_{1}}^{1}$ and $\varphi_{\bar{T}_{1}}^{1}$ have an amplitude only at the sites $a$ and $d$, while $\varphi_{T_{1}}^{2}$ and $\varphi_{\tilde{T}_{1}}^{2}$ have an amplitude only at the sites $b$ and $e$, indicating that the sites $a,d$ and $b,e$ form a sublattice as do the sites of graphene with a honeycomb structure. Because of this, the band dispersion is linear around $T_{1}$ and $\bar{T}_{1}$ and thus the Dirac cone is formed.

When the $\beta_{12}$-sheet is perturbed and is modulated to make a periodicity of $L_{x}=Ma_{x}$ and $L_{y}=Na_{y}$, in the $x$- and $y$-directions, respectively, the $\left(  M,N\right)  $ superlattice thus formed has a peculiar band structure. When $M$ is a multiple of three, the Dirac points are folded to the $\Gamma$ point of the BZ of the superlattice and the degenerate Dirac cones interfere with each other. To show the details, let us concentrate on the special case $\left(  M,N\right)  =\left(  3,2\right)  $ and assume that a perturbation sinusoidally modulates the on-site energy as
\begin{equation}
V_{n_{x},n_{y}}\left(  x,y\right)  =V_{0}\cos\left(  2\pi n_{x}\frac{x}{L_{x}%
}\right)  \cos\left(  2\pi n_{y}\frac{y}{L_{y}}\right)  ,\label{eqs4}%
\end{equation}
where $n_{x}$ and $n_{y}$ are integers. Our analysis shows that the bands are shifted in the $k_{y}$-direction when the on-site energy is perturbed as shown in Fig. S1a, which is realized when the index $\left(  n_{x},n_{y}\right)  $ is taken to be $\left(  1,6\right)  $, $\left(  2,12\right)  $, $\left( 4,12\right)  $ or $\left(  5,6\right)  $. The bands are shifted in the $k_{x}$-direction when the on-site is perturbed as shown in Fig. S1b, which is realized by taking one of $\left(  1,2\right)  $, $\left(  1,10\right)  $, $\left(  2,4\right)  $, $\left(  2,8\right)  $, $\left(  5,2\right)  $, $\left(  5,10\right)  $, $\left(  4,4\right)  $ or $\left(  4,8\right)  $. In the former case (case \textrm{i}), the original Dirac points are shifted to $\left(  0,\pm\frac{q_{y}}{L_{y}}\right)  $ with
\begin{equation}
q_{y}\simeq2\frac{V_{0}}{t_{0}}+0.0628\left(  \frac{V_{0}}{t_{0}}\right)
^{3},\label{eqqy}%
\end{equation}
thereby forming a split Dirac cone. While in the latter case (case \textrm{ii}), the original Dirac points are shifted to $\left(  \pm\frac{q_{x}}{L_{x}},0\right)  $ with
\begin{equation}
q_{x}\simeq1.23\frac{V_{0}}{t_{0}}\label{eqqx}%
\end{equation}
For other indices $\left(  n_{x},n_{y}\right)  $, the cones do no shift in the reciprocal space, while the band gap opens when $\left(  3,2\right)  ,\left( 3,10\right)  ,\left(  6,4\right)  $ or $\left(  6,8\right)  $ (or the case \textrm{iii} where the onsite-energy is distributed as shown in Fig. S1c), and does not open when $\left(  3,6\right)  $ or $\left(  6,12\right)  $ (or the case \textrm{iv}). In this way, we have seen that the splitting occurs when the sublattice symmetry is broken and that the splitting occurs in the $k_{y}$ direction ($k_{x}$ direction) when retaining (breaking) the inversion symmetry.

It is instructive to analyze an approximate solution obtained using the degenerate perturbation theory, wherein the wave function is described as a linear combination of the zero energy bands: $\varphi_{T_{1}}^{1}$,
$\varphi_{T_{1}}^{2}$,$\varphi_{\bar{T}_{1}}^{1}$, $\varphi_{\bar{T}_{1}}^{2}$. The $4\times4$ matrix $\left\langle \varphi\right\vert \mathcal{H+}V_{n_{x},n_{y}}\left\vert \varphi\right\rangle $\ at the point $\left( k_{x},k_{y}\right)  $ has the form%
\begin{equation}
\left(
\begin{array}
[c]{c|cccc}
& \varphi_{T_{1}}^{1} & \varphi_{T_{1}}^{2} & \varphi_{\bar{T}_{1}}^{1} &
\varphi_{\bar{T}_{1}}^{2}\\\hline
\varphi_{T_{1}}^{1} & A_{11} & A_{12} & C_{11} & C_{12}\\
\varphi_{T_{1}}^{2} & A_{21} & A_{22} & C_{21} & C_{22}\\
\varphi_{\bar{T}_{1}}^{1} & C_{11}^{\ast} & C_{21}^{\ast} & B_{11} & B_{12}\\
\varphi_{\bar{T}_{1}}^{2} & C_{121}^{\ast} & C_{22}^{\ast} & B_{21} & B_{22}%
\end{array}
\right)  , \label{matel0}%
\end{equation}
with%
\begin{equation}
C=t_{0}\left(
\begin{array}
[c]{cc}%
1-3e^{-iq_{y}}-4\cos\left(  q_{x}+\frac{2\pi}{3}\right)  & -8\omega\sin
^{2}\left(  \frac{1}{2}q_{x}\right) \\
8\omega^{2}\sin^{2}\left(  \frac{1}{2}q_{x}\right)  & 1-3e^{-iq_{y}}%
-4\cos\left(  q_{x}-\frac{2\pi}{3}\right)
\end{array}
\right)  , \label{matelsub1}%
\end{equation}
which can be categorized according to the cases \textrm{i} to \textrm{iv} as
\begin{equation}%
\begin{array}
[c]{c|c}%
\text{case} & \left\langle \varphi\right\vert \mathcal{H+}V_{n_{x},n_{y}%
}\left\vert \varphi\right\rangle \\\hline
\mathrm{i} & \left(
\begin{array}
[c]{cc}%
6V_{0}J & C\\
C^{\dagger} & -3V_{0}J
\end{array}
\right) \\
\mathrm{ii} & \left(
\begin{array}
[c]{cc}%
6V_{0}J & C\\
C^{\dagger} & 6V_{0}J
\end{array}
\right) \\
\mathrm{iii} & \left(
\begin{array}
[c]{cc}%
12V_{0}I & C\\
C^{\dagger} & -6V_{0}J
\end{array}
\right) \\
\mathrm{iv} & \left(
\begin{array}
[c]{cc}%
12V_{0}I & C\\
C^{\dagger} & 12V_{0}I
\end{array}
\right)
\end{array}
. \label{matel1}%
\end{equation}
where $I=\left(
\begin{array}
[c]{cc}%
1 & 0\\
0 & 1
\end{array}
\right)  $ and $J=\left(
\begin{array}
[c]{cc}%
0 & 1\\
1 & 0
\end{array}
\right)  $.

For the case \textrm{i} the eigensolutions are extremely simple. The eigenvalues around the point $\left(  0,q_{y}\right)  $ in the BZ (see Eq.\ (\ref{eqqy}) for the value of $q_{y}$), or around the split Dirac cones, are%
\begin{equation}
\varepsilon=\left\{
\begin{array}
[c]{l}%
6\left(  -V_{0}-t_{0}\sin\frac{q_{y}}{2}\right) \\
6\left(  -V_{0}+t_{0}\sin\frac{q_{y}}{2}\right) \\
6\left(  V_{0}-t_{0}\sin\frac{q_{y}}{2}\right) \\
6\left(  V_{0}+t_{0}\sin\frac{q_{y}}{2}\right)
\end{array}
\right.  \label{approxvalue}%
\end{equation}
and the corresponding eigenvectors are%
\begin{equation}
\varphi=\left\{
\begin{array}
[c]{l}%
ie^{-q_{y}/2}\left(  \varphi_{K_{1}}^{1}-\varphi_{K_{1}}^{2}\right)  +\left(
-\varphi_{\bar{K}_{1}}^{1}+\varphi_{\bar{K}_{1}}^{2}\right) \\
ie^{-q_{y}/2}\left(  -\varphi_{K_{1}}^{1}+\varphi_{K_{1}}^{2}\right)  +\left(
-\varphi_{\bar{K}_{1}}^{1}+\varphi_{\bar{K}_{1}}^{2}\right) \\
ie^{-q_{y}/2}\left(  -\varphi_{K_{1}}^{1}-\varphi_{K_{1}}^{2}\right)  +\left(
\varphi_{\bar{K}_{1}}^{1}+\varphi_{\bar{K}_{1}}^{2}\right) \\
ie^{-q_{y}/2}\left(  \varphi_{K_{1}}^{1}+\varphi_{K_{1}}^{2}\right)  +\left(
\varphi_{\bar{K}_{1}}^{1}+\varphi_{\bar{K}_{1}}^{2}\right)
\end{array}
\right.  . \label{approxvector}%
\end{equation}
The parameter dependence of the eigenvalues illustrates how the cones are split and how the bands depend on the wavenumber (linear dispersion). The eigenvectors shown in the second and the third lines of Eq.\ (\ref{approxvector}) correspond to the Dirac cone shifted in the $+k_{y}$ direction, and the first and the fourth to the cone shifted in the $-k_{y}$ direction. These solutions are made by a simple superposition of the unperturbed states and are allowed to propagate in the $y$ direction. The simplicity reflects the fact that the sublattice site pairs $a,d$ versus $e,b$ are equivalently perturbed by $V_{1,6}\left(  x,y\right)  $ (see Fig. S1a).

The splitting of the Dirac cone can be similarly observed from the first-principles calculations, as shown in Fig. S2b. Indeed, when perturbed with $V_{3,1}$, the Dirac cone is split in the $y$ direction as shown in Fig. 2g. \begin{table}[ptb]%
\begin{tabular}
[c]{c|ccccc}%
$\tilde{t}\backslash t$ & $a$ & $b$ & $c$ & $d$ & $e$\\\hline
$a$ & $0.500\text{ eV}$ & $-2.04\text{ eV}$ & $-1.79\text{ eV}$ & $-$ &
$-2.12\text{ eV}$\\
$b$ & $-$ & $0.532\text{ eV}$ & $-1.84\text{ eV}$ & $-1.91\text{ eV}$ & $-$\\
$c$ & $0.308,\text{ }0.389\text{ eV}$ & $-0.130\text{ eV}$ & $0.310,\text{
}-0.126\text{ eV}$ & $=t_{bc}$ & $=t_{ac}$\\
$d$ & $0.421\text{ eV}$ & $-$ & $=\tilde{t}_{bc}$ & $=\tilde{t}_{bb}$ &
$=t_{ab}$\\
$e$ & $-0.143\text{ eV}$ & $0.421\text{ eV}$ & $=\tilde{t}_{ac}$ & $-$ &
$=\tilde{t}_{aa}$%
\end{tabular}
\caption{The transfer of the MLWF effective Hamiltonian shown in a matrix
form. Those having magnitude larger than 0.1 eV are shown. Those for the
nearest neighbor site, $t_{ij}$, are shown in the upper triangular part and
others, $\tilde{t}_{ij}$, are shown in the lower triangular part. Two values
are assinged to $\tilde{t}_{ac}$ and $\tilde{t}_{cc}$ because there are two
inequivalent pairs: The former value corresponds to those pairs with common
$x$-coordinate and the latter to those with different $x$-coordinate.}%
\label{table1}%
\end{table}

\begin{figure}[htbp]
\includegraphics[width=10 cm]{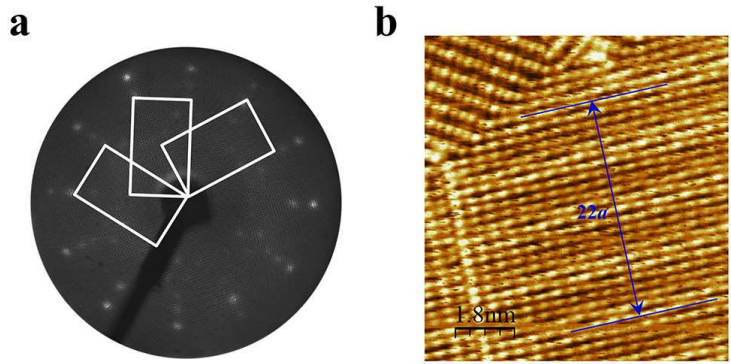}
\caption {The applied on-site perturbation. The black and red circles correspond to positive and negative value, and the large circles correspond to the perturbation of magnitude V$_0$, the medium to V$_0$/2, and the small to V$_0$/4.}
\end{figure}

\begin{figure}[htbp]
\includegraphics[width=10 cm]{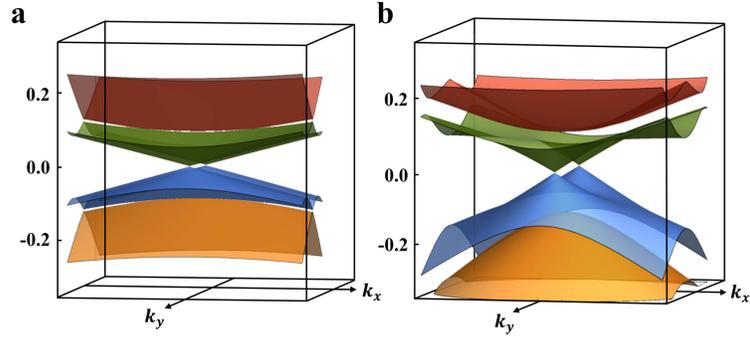}
\caption {Band structures of the (3, 1) superlattice perturbed with {\it V$_{1,3}$(x,y)} from the TB model (a) and first-principles calculations (b). Only the central region in the BZ is shown here: -0.1$\pi$/$L_x$ $\leq$ $k_x$ $\leq$ -0.1$\pi$/$L_x$, -0.1$\pi$/$L_y$ $\leq$ $k_y$ $\leq$ -0.1$\pi$/$L_y$, where $L_x$ and $L_y$ are the lattice constants of the extended cell.}
\end{figure}

\bigskip\leftline{\bf 2. Substrate-induced shift of the Dirac cones}

\bigskip To investigate the role of metal substrates on the energy shift of the Dirac cones, we performed first-principles calculations using a simplified overlayer model without moir\'{e} patterns [the moir\'{e} pattern has been taken into account in Figs. 3(h)-3(k)]. The structure model was constructed by placing a 1$\times$1 lattice of the $\beta_{12}$-sheet on a Ag(111)-$3\times\sqrt{3}$ surface slab consisting of six Ag layers. The lattice constants of the $\beta_{12}$-sheet have been adjusted to fit the periodicity of Ag(111). This simplified structure model, obtained without performing geometry optimization, has the lowest energy when the $\beta_{12}$-sheet is 2.54 {\AA} away from Ag(111) and the boron atoms are located at the bridge sites [Figs. S3(a) and S3(b)]. The calculated band structures in Fig. S3c show that the Dirac point shifts 1.0 eV downward compared with that of the free-standing $\beta_{12}$-sheet, as indicated by the red and black circles. However, the wave function of the Dirac cones has almost the same spatial distribution as that of the free-standing $\beta_{12}$-sheet (yellow circles). This fact indicates that Ag(111) substrate has little effect on the nature of the Dirac cones, and only shift the Dirac cone downward. The Dirac cone is further shifted downward with the decrease of the distance between the $\beta_{12}$-sheet and Ag(111), and finally crosses E$_F$ when the distance is 1.5 {\AA}.

To reveal the origin of the energy shift of the Dirac cones, we further examined the electronic structure of the $\beta_{12}$-sheet. The electrons are found to be transferred from Ag(111) to the boron sheet, and, on average, each B atom obtains 0.06 electron according to our population analysis. However, we found that the energy shift of the bands does not have a strong dependence on the amount of charge transferred. This fact indicates that the charge transfer may not be the only reason for the band shift, and the orbital hybridization between B and Ag may also play a significant role. The charge transfer and orbital hybridization do not affect the main nature of the Dirac cones in our simplified structure model. After considering the moir\'{e} pattern, the hybridization between B and Ag will split the Dirac cones. It should be noted that the energy shift from our simplified structure model is not precise, but it provides strong evidence for the downward shift of the Dirac cone when it is placed on a Ag(111) substrate.

\begin{figure}[htbp]
\includegraphics[width=14 cm]{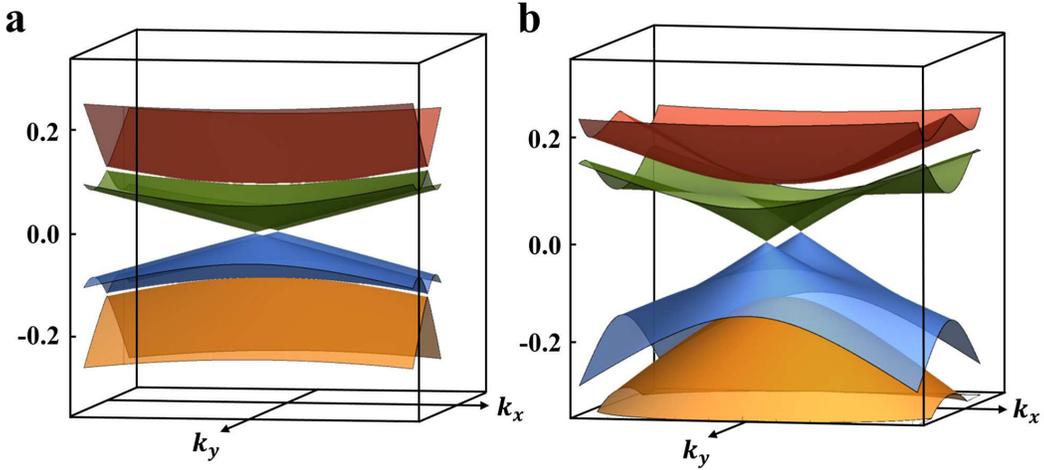}
\caption {(a) and (b) The $\beta_{12}$-sheet placed on Ag(111) surface. The green and grey balls indicate the B and Ag atoms, respectively. The blue rectangles indicate the unit cells. The yellow circles represent the spatial distribution of the squared wave function that corresponds to the Dirac cones. (a) and (b) corresponds to different states of the Dirac cones. (c) Band structures of the free-standing $\beta_{12}$-sheet (red dotted lines) and the $\beta_{12}$-sheet placed Ag(111) (grey solid lines). The Dirac cones are indicated by the red and black circles. The red and grey arrows indicate other typical bands that shift downward when the boron sheet is placed on a Ag(111) substrate.}
\end{figure}

\bigskip\leftline{\bf 3 Experimental methods}

\bigskip The samples were prepared and measured in a combined system equipped with a preparation chamber and analysis chamber at BL-2A MUSASHI beamline of Photon Factory, KEK. The base pressures in the preparation chamber and ARPES chamber were 1$\times$10$^{-10}$ mbar and 7$\times$10$^{-11}$ mbar, respectively. Single crystal Ag(111) was cleaned by repeated sputtering and annealing cycles and the cleanness of Ag(111) was confirmed by the sharp LEED patterns and Shockley surface states. Boron was evaporated from a commercial e-beam evaporator at a flux of ~0.03 ML/min, while keeping the Ag(111) at approximately 600 K. The LEED pattern of the as-prepared sample is shown in Fig. S4. The sample was transferred to the analysis chamber for ARPES measurements without breaking the vacuum. All the ARPES measurements were performed at 20 K using a Scienta SES2002 analyzer. We used linearly polarized light ({\it s} and {\it p} polarization) and the photon energy could be tuned from the ultra-violet to soft X-ray regions.

The scanning tunneling microscopy (STM) measurements were performed in a home-built MBE-STM system. The samples were prepared in the preparation chamber using the same method as described above. After preparation, the sample was immediately transferred to the STM chamber for measurements. All STM images were acquired in the constant current mode.

\bigskip\leftline{\bf 4. The moir\'{e} pattern of the $\beta_{12}$-sheet on Ag(111)}

\bigskip Fig. S4a shows the LEED pattern of the $\beta_{12}$-sheet on Ag(111). Besides the primary spots of the $\beta_{12}$-sheet, there are weaker spots that correspond to the moir\'{e} pattern. In our STM experiments, we also observe a parallel striped pattern, which is incommensurate to the lattice of the $\beta_{12}$ sheet (Fig. S4b). The periodicity of the striped pattern is approximately 1.6 nm, which is around 5.5 times the lattice constant of the $\beta_{12}$-sheet. The moir\'{e} pattern imposes a periodic perturbation to the boron lattice, resulting in the splitting of the Dirac cones, as discussed in the text.

\begin{figure}[htbp]
\includegraphics[width=13 cm]{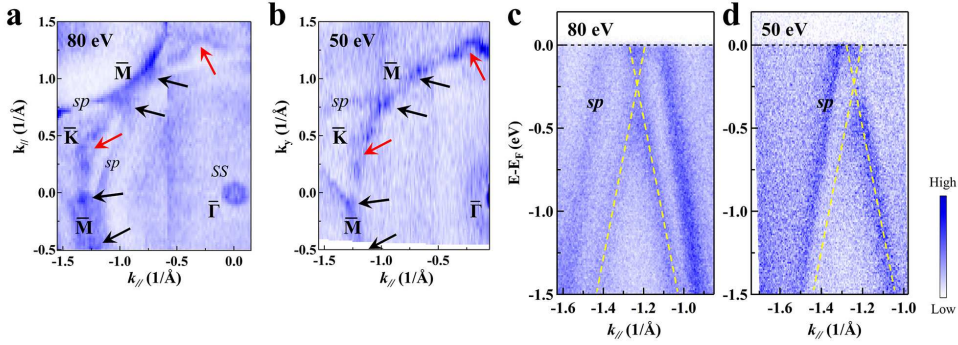}
\caption {(a) LEED pattern of boron on Ag(111). The rectangles indicate three equivalent domains with different orientations. (b) High resolution STM image of the $\beta_{12}$-sheet. The rectangular lattice and moir\'{e} pattern are clearly observed.}
\end{figure}

\bigskip\leftline{\bf 5. Band structures measured with different photon energies}

\bigskip To confirm the two-dimensional character of the boron bands, we performed ARPES measurements with different photon energies, as shown in Fig. S5. The momentum positions of the Dirac cones are the same using photon energies of 80 eV and 50 eV, in agreement with the surface origin of the boron bands. It should be noted that only one branch of the Dirac cone is visible in Fig. S5 owing to the matrix element effects, similar to the case for graphene\cite{Bostwich2007,Gierz2008}.

From Fig. S5, one can also observe a gap-like feature at the Dirac point. This ``gap" is gradually filled with an increasing intensity of the left branch of the Dirac cone (referenced from the lower Dirac cone). The ``gap" is still visible when measured with 80-eV photons (Fig. 3d) because the intensity of the left branch is weaker than that of the right branch. In graphene, there is a ``kink'' at the Dirac point that originates from the many-body interactions\cite{Bostwich2007}, which resembles the gap-like feature in the $\beta_{12}$ boron sheet. To confirm the origin of this ``gap'' or ``kink'' feature, further theoretical and experimental studies are necessary.

\begin{figure}[htbp]
\includegraphics[width=15 cm]{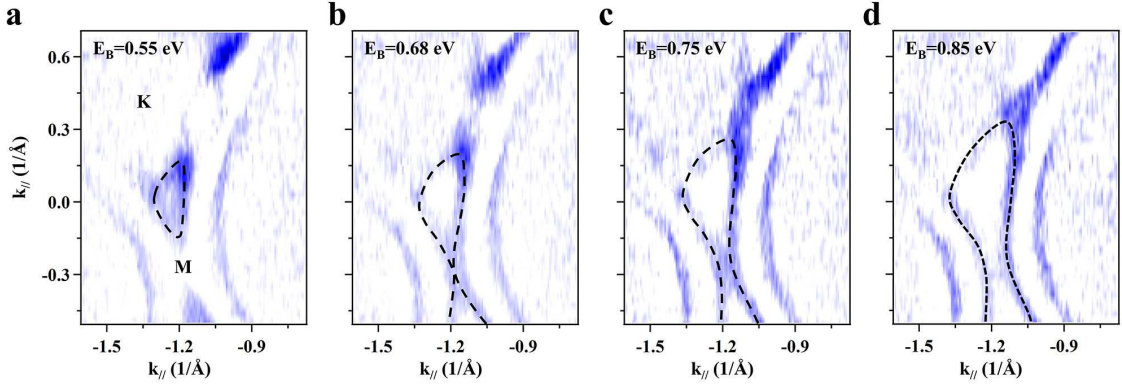}
\caption {(a) and (b) Fermi surfaces measured using photon energies of 80 eV and 50 eV, respectively. The black and red arrows indicate the bands of the boron layer. (c) and (d) ARPES intensity plots along cut 1 (Fig. 3a) measured with photon energies of 80 eV and 50 eV, respectively. The yellow dashed lines indicate the Dirac cones.}
\end{figure}

\bigskip\leftline{\bf 6. First-principles calculations on the B/Ag(111) system}

\bigskip First-principles calculations, using density functional theory (DFT), were performed with the Vienna Ab initio Simulation Package (VASP)\cite{Krasse1996}. The projector augmented-waves (PAW) method\cite{Blochl1994} and Perdew-Burke-Ernzerhof (PBE) exchange-correlation\cite{Krasse1996} were used. The plane-wave cutoff energy was set to 400 eV and the vacuum space was set to be larger than 15 {\AA}. The unit cell in our calculation contained a $\beta_{12}$-sheet on a four-layer 3$\times$3$\sqrt3$ Ag(111) surface. The two-dimensional BZ was sampled using the Monkhorst-Pack scheme\cite{Monkhorstl1976}. A k-point mesh 5$\times$5$\times$1 was used for the structural optimization, while a 6$\times$6$\times$1 mesh was adopted in the self-consistent calculations. Using the conjugate gradient method, the positions of the atoms are optimized until the residual force on each atom is less than 0.05 eV/{\AA}. The calculated band structure was further processed by the orbital-selective band unfolding technique\cite{Medeiros2014,Medeiros2015}. Owing to the high surface-sensitivity of ARPES, the calculated band structure of boron/Ag(111) is projected onto the boron layer for comparison with the ARPES spectra.

Because the PBE function usually overestimates the chemical bond length, the lattice constant of Ag(111) used in the calculation process was 3\% larger than the experimental value. As a result, compared with the freestanding $\beta_{12}$-sheet, the lattice constants are 5\% smaller and 2\% larger in the a$_x$ and a$_y$ directions (Fig. 1b).

\end{document}